\documentclass[twocolumn,fleqn]{article}
\begin{document}
\twocolumn[
{\bf
RELATIONS DEPENDENT ON NEW FUNDAMENTAL CONSTANTS  \\ 
AMONG SPACETIME OBSERVABLES OF QUANTUM PARTICLE }

\bigskip

{\large\bf V.V. Khruschov$^{1,2}$}
 
\medskip

{\it $^{1}$Center for Gravitation and Fundamental Metrology, VNIIMS, 46 Ozyornaya St.,
Moscow 119361 
\\ 
$^{2}$National Research Centre Kurchatov Institute, Ac. Kurchatov Sq. 1, Moscow 123182, Russia 
}

\medskip

%{\it Received 30 June 2009}

\vspace*{-0.5cm}

%\texttt{ 
\begin{center}
\begin{abstract}
\parbox{15cm}{Generators of spacetime translations and  Lorentz group transformations form the Lie algebra of the Poincar\'e group and give rise to the Casimir invariants for a  specification of elementary particle characteristics. 
Moreover quantum operators of coordinate and momentum components  of a particle in  Minkowski spacetime together with Lorentz group generators belong to the known
noncommutative algebra. This algebra can be generalized under some constraints, in particular, the Lorentz invariance condition. The generalized algebra depends on the new fundamental constants with dimensions of length (L), mass (M) and action (H). Quantum fields, which can be constructed with the help of representations of this algebra, are referred to as HLM generalized quantum fields and the associated  particles as HLM quantum particles. Relations 
between spacetime observables of a HLM quantum particle depend on the new constants and lead to complication of the Poincar\'e invariant equations for canonical  fields. The modification of a quantum measurement procedure is needed in order to take into account inherent features of HLM quantum particles. 

\smallskip
%\medskip


%\smallskip
\noindent{\bf DOI:} 10.1134/S0202289309040069
}

\end{abstract}
\end{center}
\bigskip
\smallskip
]

\noindent\textbf{\large 1. Introduction}

\smallskip

 It is well known that the Poincar\'e symmetry, which is the spacetime 
symmetry of the orthodox relativistic quantum field theory (QFT),
originates from the isotropy and  homogeneity of Minkowski spacetime and is 
based on observations of macro- and micro-phenomena concerning 
observable physical bodies and particles. Today the Poincar\'e symmetry is a basis of any theory of fundamental interactions
for quantum particles, for instance, the Standard Model of weak, electromagnetic and strong interactions (SM). 
In particular, the quantum chromodynamics (QCD), which is the QFT for quarks  and is the part of the SM, incorporates the Poincar\'e group as the group of spacetime symmetries. The QCD describes well numerous experimental data at high energies where the perturbation theory works \cite{yndu}.  But the confinement of quarks is outside of the QCD perturbation theory and may be the QCD itself. So it is desirably to generalize the Poincar\'e group of spacetime symmetries for description of quark properties taking into account that quarks have not been observed as free particles in the Minkowski spacetime. In general case it is good reason to consider various  generalizations of the Poincar\'e symmetry bearing in mind complicated features of known and possibly  unknown interactions of elementary particles.

  Studies along 
these lines have been carried out  in the context  of  theories with new fundamental physical constants additional to
the well known  $c$ and $\hbar$. Beginning with  Snyder's 
work \cite{sny}, the theory with a fundamental length has been elaborated 
\cite{gol,kad}. However, in a theory of this sort, the reciprocity between
coordinate and momenta,  proposed by Born  \cite{born},
was broken. The reciprocity  was restored in by Yang \cite{yang}, 
who added a fundamental mass  to the modified theory (see also \cite{lez-beg}). More general algebras depended on the constants with the dimensions of length (L),  mass (M),  and action (H) were considered in the works \cite{lezkh}.

 In the present paper some characteristics of space-time symmetries for quantum particles that generalize Poincar\'e group characteristics are found. As a example of such particles we use quarks, which have some inexplicable so far properties.  We investigate the general Lie algebra
for quantum operators of coordinate, momentum and angular momentum  
 of a quantum particle that in a limiting case is reduced to the Lie algebra of observables of the canonical QFT. Quantum fields, which can be constructed with the help of representations of the general algebra, are referred to as HLM generalized quantum fields and the associated  particles as HLM quantum particles. We present relations 
between spacetime observables of a HLM quantum particle, which depend on the new 
constants L, M and H. Some of these relations lead to the equations, which generalize the Poincar\'e invariant equations for canonical  scalar and spinor fields. Operators of a momentum and coordinate of a HLM quantum particle are noncommutative that result in significant modification of a quantum measurement procedure. Lower limits to products of dispersions of 
 quantum particle observables are found  taking into account HLM generalized commutation relations.  
  
 The  paper  is organized as follows:  Section 2 is
devoted to the method for obtaining HLM generalized commutation relations which lead to a Lie algebra of spacetime observables. The equation for scalar field invariant with respect to transformations generated by the HLM algebra is presented in  Section 3. In  Section 4 it has been suggested to apply the generalized symmetries to 
description of quarks. The equations for generalized spinor fields are given 
 in  Section 5. In  Section 6  the lower limits to products of dispersions of spacetime observables are presented which depend on new constants. The important results of this paper, which are of interest for description of HLM particles and their search, are pointed out in  Section 7. 

\smallskip

\noindent{\large\bf 2. Lie algebra of spacetime observables for HLM generalized quantum fields}

\smallskip

     We consider a Lie algebra $g$ of observables for HLM  generalized quantum fields,
when coordinate and momenta are noncommutative in general case and form a generalized quantum phase space. 
%The algebra $g$  generated by the  observables can 
%depend on extra fundamental constants other than the well 
%known $c$ and $\hbar$ \cite{lez1, khru1}. 
To restrict  the considerable list
of  symmetries corresponding to the algebra $g$,
the following natural conditions are imposed:
  a) the generalized algebra $g$ of observables should be a  Lie algebra;
    b) the  dimension of $g$ should coincide with the dimension of
the algebra of canonical quantum theory observables in the Minkowski spacetime;
    c) physical dimensions of the observables, representing the 
generators of $g$, should be the same as the canonical ones;
    d) the algebra $g$ should contain the Lorentz algebra $l$ as its
subalgebra and  commutation relations of the  generators of $l$
with other generators, should be the same as the canonical ones.

Under these conditions, the most general algebra has been derived \cite{lezkh},
which depends on new constants with the dimensions of length, mass and action,
and the commutation relations can be written as ($i, j, k, l = 0, 1, 2, 3$)
\[
[F_{ij}, F_{kl}]=if(g_{jk}F_{il}-g_{ik}F_{jl}+g_{il}F_{jk}-g_{jl}F_{ik}),
\]
\[
[F_{ij}, p_{k}]=if(g_{jk}p_{i} - g_{ik}p_j), 
\]
\[ [F_{ij}, x_k]=if(g_{jk}x_i - g_{ik}x_j),
\]
\[
[F_{ij}, I]=0, \quad [p_i, p_j ]=(if/L^2)F_{ij}, 
\]
\begin{equation}
[x_i, x_j]=(if/M^2)F_{ij}, 
\label{ee1}
\end{equation}
\[ [p_i, x_j]=if(g_{ij}I +  F_{ij}/H),
\]
\[
[p_i, I]=if(x_i/L^2 - p_i/H),  
\]
\[ [x_i, I]=if(x_i/H - p_i/M^2).
\]

The first relation specifies the algebra $l$, while 
the second, third and fourth relations specify the tensor
character for the well-known physical quantities. The fifth and sixth
relations lead to noncommutativity of components as $p$ and $x$. 
The seventh, eighth and ninth relations  generalize
 the Heisenberg relations. The system of relations (1) is written 
in units with $c = 1$ ($c$ is the velocity of light), 
it contains four dimensional parameters: $f$(action), $M$(mass), $L$(lenght), 
and $H$(action). But in the limiting case  $M \to\infty$, $L\to\infty$, 
$H\to\infty$, the system (1) should transform to the system of relations
for the canonical quantum theory, so $f = \hbar$. More generally, 
$f= f(M,L,H)$ and in the limiting case $f(M,L,H)\to\hbar$.

The algebra (\ref{ee1}) contains as special cases diverse
Lie algebras of generalized space-time symmetries. The  condition 
for the algebra (\ref{ee1}) to be  semisimple  can be presented in the
 form: $(M^2L^2 - H^2)/M^2L^2H^2\ne 0$. If this condition is fulfilled,
$g$ is isomorphic to a pseudoorthogonal algebra $o(p,q)$, $p+q=6$ depending on 
selected $M$, $L$ and $H$ values  ($M$ and $L$ can be real and pure imaginary).  In other cases it is isomorphic to a direct or a semidirect product of a pseudoorthogonal algebra and an Abelian
or integrable algebra \cite{lezkh}. 
%The ranges of
%the  M, L and H parameters, which are matched to the o(3,3),
%o(4,2), and o(5,1) algebras, are given below. If $sign(H^2- M^2L^2)=-1$, 
%$sign(M^2)=sign(L^2)$, or $sign(H^2- M^2L^2)=1$, 
%$sign(M^2)=-sign(L^2)$, then $g=o(2,4)$.
%If $sign(H^2- M^2L^2)=1$, 
%$sign(M^2)=sign(L^2)=1$, then $g=o(1,5)$, if $sign(H^2- M^2L^2)=1$, 
%$sign(M^2)=sign(L^2)=-1$, then $g=o(3,3)$. 

Any one of the algebras (\ref{ee1}) can be a basis for a new field theory. Some possible applications of the algebras (\ref{ee1}) at different scales are indicated in the works \cite{tol,kh,lez}.  The central problem  is finding a really suitable model for description of a new phenomenon in our world using appropriate one of these algebras. For instance, in Ref. \cite{lez} the properties caused of the commutation relations (\ref{ee1}) attribute to the spacetime manifold itself. The following sections of the present paper are related to a possible application of the algebras (\ref{ee1}) for description of properties of  hypothetical quantum particles. Equations for scalar and spinor particles are given, in the last case it is thought of especially quarks.  

%Irreducible  representations for the pseudoorthogonal algebras 
%of rank 3 are determined with the help of eigenvalues of  three Casimir
% operators:
%$K_1=\varepsilon_{IJKLMN}F^{IJ}F^{KL}F^{MN}$,
% $K_2=F_{IJ}F^{IJ}$, $K_3=(\varepsilon_{IJKLMN}F^{KL}F^{MN})^2$, 
%$I,J,K,L,M,N=0,1,2,3,4,5$.
%For instance, the second-order invariant operator $K_2$  in terms
%of $I, p, x$, and $F$ can be written in the form:
%\[
%C_2 = \sum_{i<j}F_{ij}F^{ij}(1/M^2L^2 - 1/H^2) + I^2 +
%\]
%\begin{equation}
% (x_ip^i + p_ix^i)/H - x_ix^i/L^2 - p_ip^i/M^2.
%\end{equation}

%The mathematical properties of the generalized algebra (\ref{al1}), 
% have been studied in   \cite{lez1,khru1,mendes,chry}.
%Apart  from its mathematical properties, the  algebra $g$ is an object of 
%interest in modern physical applications as well. For instance, in 
% \cite{lez2}, it is suggested  to apply the algebra 
%(\ref{al1}) in  classical physics on astronomical scales.

\smallskip

\noindent{\large\bf 3.  Equation for HLM generalized quantum scalar field
}

Representations of the HLM algebra are characterised with values of three Casimir
 operators:
$C_1=\varepsilon_{IJKLMN}F^{IJ}F^{KL}F^{MN}$,
 $C_2=F_{IJ}F^{IJ}$, $C_3=(\varepsilon_{IJKLMN}F^{KL}F^{MN})^2$, 
$I,J,K,L,M,N=0,1,2,3,4,5$. Then, for example,  a HLM generalized quantum scalar field obeys an equation of the second order invariant with respect to transformations generated by HLM algebra operators (\ref{ee1}) in a $\xi$-representation \cite{khlez}:
\[
 (\sum_{i<j}F_{ij}F^{ij}(1/M^2L^2 - 1/H^2) + I^2 +   
\]
\begin{equation}
(x_ip^i + p_ix^i)/H - x_ix^i/L^2 - p_ip^i/M^2)\Phi(\xi)=0 
 \label{sfe}
\end{equation} 

An interesting case is when the $p_i$ and $x_j$ operators are commutative in a similar way it is in the standard theory framework and the only one new constant $H$ remains. Then the H algebra constitutes from the following operators  in the $\xi$-representation \cite{kh-94}:
\[
p_i= i\hbar\frac{\partial}{\partial\xi^i}, \, x_i= i\hbar(a\xi_i+\frac{\xi_i\xi^m}{H}\frac{\partial}{\partial\xi^m}-\frac{\xi^2}{2H}\frac{\partial}{\partial\xi^i})
\]
\begin{equation}
 F_{ij}= i\hbar(\xi_i\frac{\partial}{\partial\xi^j}-\xi_j\frac{\partial}{\partial\xi^i}), I=i\hbar(a + \frac{\xi^m}{H}\frac{\partial}{\partial\xi^m}),
 \label{sfo}
\end{equation} 
\noindent where $a$ is an arbitrary real number.

\smallskip

\noindent{\large\bf 4.  The generalization of spacetime symmetries for quarks
}

\smallskip

Below we apply the algebra (\ref{ee1}) for description of quarks and find some new features which arise in quark spacetime properties in this case. First of all quarks ought be characterized with new invariants instead of the Poincare algebra invariants connected with mass and spin. So a generalized quark field  will be given with values of three Casimir
 operators: $C_1$, $C_2$ and $C_3$. 

In the presence of the constant $H$  the CP-invariance does not hold for the algebras (\ref{ee1})  \cite{lezkh}. So if we take into account the $CP-$invariance of strong interactions we must put $H = \infty$ for quarks and obtain  the following 
nonzero commutation relations for $p_i$, $x_j$ and $I$ (below we use the natural units $c=\hbar=1$):
\[
 [p_i, p_j ]=(i/L^2)F_{ij},
\]
\[ 
 [x_i, x_j]=(i/M^2)F_{ij},  
\]
\begin{equation}
 [p_i, x_j]=ig_{ij}I,
\label{al2}
\end{equation}
\[
[p_i, I]=i(x_i/L^2), 
\]  
\[ 
[x_i, I]=i(- p_i/M^2).
\]
%\noindent where $\hbar=1$.

Let us name the quantum fields which can be formed with the help of an irreducible
representation of the LM algebra (\ref{al2}) as the 
LM generalized quantum fields.  Using of the Casimir operator of the second order $C_2$ we suppose that LM field obeys the following equation \cite{kh-97}:
\[
 (\sum_{i<j}F_{ij}F^{ij}(1/M^2L^2) + I^2 -   
\]
\begin{equation}
 - x_ix^i/L^2 - p_ip^i/M^2)\Phi(\xi)=0 
 \label{lme}
\end{equation} 

\smallskip 

\noindent{\large\bf 5.  Equations for LM generalized quantum spinor fields 
}

\smallskip

When a LM generalized quantum field have a nonzero spin value there are other equations in addition to the equation (\ref{lme}). Two equations can be given for spinor fields. The equation (\ref{nde}) is not invariant with respect to the space parity (P), while the equation (\ref{ide}) is invariant with respect to this discrete transformation.

\[
 (\gamma_ip^i - \gamma_i\gamma_5x^i\zeta_1\zeta_2\sqrt{-\frac{M^2}{L^2}}- \gamma_5I\zeta_2\sqrt{-M^2}  
\]
\begin{equation}
 -\sum_{i<j}\gamma_i\gamma_jF^{ij}\frac{\zeta_1}{\sqrt{L^2}}-n)\psi(\xi)=0. 
 \label{nde}
\end{equation} 
\[
 (\sigma_0\otimes\gamma_ip^i - \sigma_3\otimes\gamma_i\gamma_5x^i\zeta_1\zeta_2\sqrt{-\frac{M^2}{L^2}}
\]
\begin{equation}
  -\sigma_3\otimes\gamma_5I\zeta_2\sqrt{-M^2}
\label{ide}
\end{equation}  
\[
 -\sigma_0\otimes\sum_{i<j}\gamma_i\gamma_jF^{ij}\frac{\zeta_1}{\sqrt{L^2}}-\sigma_0\otimes n)\Psi(\xi)=0. 
\] 
\noindent where $F^{ij}$ are Lorentz generators, 
%=L^{ij}+S^{ij}$, $L^{ij}$ is a orbital part of a 
%$F^{ij}$, $S^{ij}$ is a spinorial part of a Lorentz generator $F^{ij}$, 
$n$ is an arbitrary number, $\sigma_0$ and $\sigma_3$ are Pauli matrices,
$\gamma_i$, $i=0,1,2,3$ are Dirac matrices, $\zeta_1=\pm 1$, $\zeta_2=\pm 1$.

The coefficients $\Xi$, $\Delta$ and $\Sigma$ of the equation (5) from Ref. \cite{khru2}
are made more precise with the coefficients of the equations (\ref{nde}) and (\ref{ide}).
 The properties  of  solutions 
 of Eqs. (\ref{nde}) and (\ref{ide}) are largely dependent on values of the  coefficients of these equations and should be studied closely in the each case. 
% In the limiting case  $L\to\infty$, $M\to\infty$ these equations    reduce to the Dirac %equation  of the canonical quantum theory. 

\smallskip

\noindent\textbf{\large 6. Dispersions for noncommutative observables 
of a LM quantum particle }

\smallskip

In this section a brief description of modification of a quantum measurement procedure for quark spacetime observables is outlined. The especial feature consists in appearance of corrections dependent on the constants L and M in commutation relations for observables of the canonical quantum theory. The main characteristic, that must be determined in this case, is a dispersion of the spacetime observable entered in the LM algebra (\ref{al2}). The dispersion of a observable A in a state $\Psi$ is
\begin{equation}
D(A)_{\Psi}= (\Delta A_{\Psi})^2=<A^2>_{\Psi}-<A>^2_{\Psi},
\label{dis}
\end{equation}
\noindent where $\Delta A_{\Psi}$ is a standard uncertainty of a measurement value $<A>_{\Psi}$ in a state $\Psi$ (see, e.g. \cite{sen}).

Taking into account the commutation relations (\ref{ee1}) at $H=\infty$ we obtain the lower limits to products of dispersions of quark spacetime observables, which depend on the new constants $L$ and $M$.

\[
 D(p_i)_{\Psi}D(p_j)_{\Psi}\ge\left(\hbar^2/4\right)\left|\langle
F_{ij}/L^2\rangle_{\Psi}\right|^2,
\]
\[
D(x_i)_{\Psi}D(x_j)_{\Psi}\ge\left(\hbar^2/4\right)\left|\langle
F_{ij}/M^2\rangle_{\Psi}\right|^2,\\\nonumber
\]
\[
 D(p_i)_{\Psi}D(I)_{\Psi}\ge\left(\hbar^2/4\right)\left|\langle
 x_{i}/L^2\rangle_{\Psi}\right|^2,\\\nonumber
\]
\begin{equation}
 D(x_i)_{\Psi}D(I)_{\Psi}\ge\left(\hbar^2/4\right)\left|\langle
 p_{i}/M^2\rangle_{\Psi}\right|^2
\label{disne}
\end{equation}
\smallskip

\noindent\textbf{\large 7. Conclusions  }

\smallskip

We have considered in some detail the general HLM
algebra of the physical observables which depends on 
additional constants with the dimensions of length (L), mass (M) and action (H). 
It has been supposed that the representations of the  HLM
algebra (\ref{ee1}) can be used for construction of quantum fields which describe hypothetical 
HLM particles. In particular the possibility has been considered that the strongly interacting 
fundamental particles (quarks) can be particles of such type.

Some relations 
between spacetime observables of a HLM quantum particle depended  on the new 
constants has been presented including the equations for scalar and spinor fields 
(\ref{sfe}), (\ref{lme}), (\ref{nde}) and (\ref{ide}). 
The modification of a quantum measurement procedure relates to the nonzero 
lower limits to products of dispersions for 
spacetime observables (\ref{disne}). These characteristics take into account inherent 
features of HLM quantum particles and can be used for their search.

     In  summary,  it  can  be  said that a study  of the
generalized  algebra  (\ref{ee1})  and  the properties  of  solutions 
 of Eqs. (\ref{sfe}), (\ref{lme}), (\ref{nde}) and (\ref{ide}) are important objectives for 
achievement of  mathematical completeness of the HLM generalized quantum fields approach. A  further 
investigation in this direction  is in progress  now.
%  and will be presented  elsewhere.

\smallskip

%\noindent\textbf{\large Acknowledgement}

%\smallskip

%The author is grateful to V.N. Melnikov, K.A. Bronnikov, V.D. Ivashchuk
%and S.V. Bolokhov for useful discussions.

\end{document}